# Photovoltaic cell based on *n*-ZnO microrods and *p*-GaN film

B. Turko[1*], V. Vasil'yev[1], B. Sadovyi[1, 2], V. Kapustianyk[1], Y. Eliyashevskyi[1], R. Serkiz[1]

[1]*Ivan Franko National University of Lviv, Dragomanova Str., 50, Lviv, 79005, Ukraine,*

[2]*Institute of High Pressure Physics PAS, 29/37, Sokolowska Str., 01-142, Warsaw, Poland*

**Corresponding author:** Vladyslav Vasil'yev (tyrko_borys@ukr.net, phone: +38 032 239-46-47)

**Abstract**

The photovoltaic cell based on *p*-GaN film/*n*-ZnO microrods quasiarray heterojunction was fabricated. According to the scanning electron microscopy data, the ZnO array consisted of the tightly packed vertical microrods with a diameter of approximately 2–3 μm. The turn-on voltage of the heterojunction of ZnO/GaN (rods/film) was around 0.6 V. The diode-ideality factor was estimated to be of around 4. The current-voltage characteristic of the photovoltaic cell under UV LED illumination showed an open-circuit voltage of 0.26 V, a short-circuit current of 0.124 nA, and a fill factor of 39 %, resulting in an overall efficiency of $1.4 \cdot 10^{-5}$ %. These results may be useful in the engineering of electronic devices based on the materials with optical transparency.

## 1 Introduction

Power generation by fossil-fuel resources has peaked, whilst solar energy is predicted to be at the vanguard of energy generation in the near future. Moreover, it is predicted that by 2050, the generation of solar energy will have increased to 48% due to economic and industrial growth [1]. Photovoltaic cells are usually designed and fabricated to utilize visible light, which is the major part of solar energy, to generate electricity. Moreover, the fabricated transparent photovoltaic cells transforming the ultraviolet (UV) light into electric energy would be used as window glass of buildings or cars. Then the transparence of the glass will not be altered much, and may reduce the possible harm caused by UV irradiation to humans. The generated electricity may also be used to power household appliances [2].

ZnO/GaN heterostructure-based light-emitting devices, photodetectors and lasers have already been demonstrated [2–4], but there are few reports on the photovoltaic cells based on such structures [2, 5]. The reported photovoltaic cells [2, 5] were made on the basis of *p*-GaN epitaxial films and ZnO films grown by the molecular beam epitaxy [2] or obtained by the radio-frequency magnetron sputtering [5]. One can suppose that the efficiency of the mentioned solar cells would be increased due to application of the array of ZnO nano- or microelements in the above mentioned systems due to considerable growth of the effective surface area of such a heterostructure.

In this paper, we report the fabrication and characterization of the photovoltaic cell based on *p*-GaN film/*n*-ZnO microrods array heterojunction.

## 2 Experimental

The investigated heterojunctions were fabricated on the basis of *p*-type GaN templates purchased from UNIPRESS from Poland such as magnesium-doped GaN (0001)-oriented 2 μm thick layer grown by metalorganic vapor phase epitaxy (MOVPE) on 430 μm thick sapphire substrate with GaN non-conductive 1.5 μm thick buffer layer. According to the passport data *p*-GaN film is characterized by a relatively low dislocation

density – $(3–5)\times10^8$cm$^{-2}$; the concentration of the embedded magnesium impurity is $2\times10^{19}$cm$^{-3}$; the concentration of the electrically active holes is $(2–3)\times10^{17}$cm$^{-3}$.

ZnO hexagonal microrods (Fig. 1) were grown by the method of gas-transport reactions on *p*-GaN epitaxial film [4].

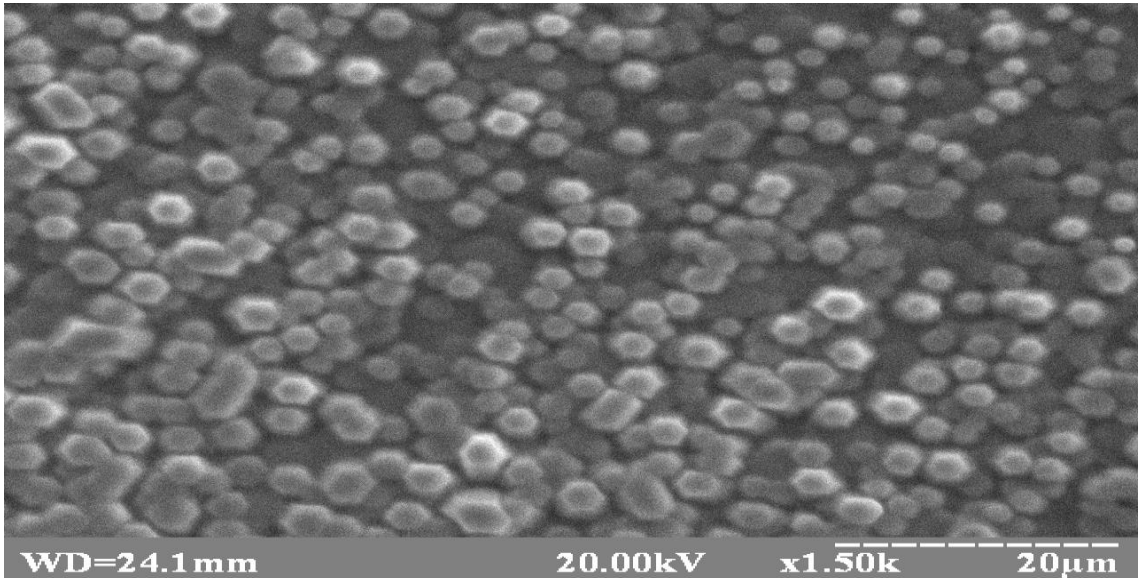

**Fig. 1** Microphotographs of ZnO hexagonal microrods grown by the method of gas-transport reactions

The contacts on *p*-GaN tamplate were deposited using thermal evaporation of Ni (30 nm) followed by Au (35 nm) [3, 4]. The quartz-crystal microbalance was used as a thin film deposition monitor. To produce the photovoltaic cell, the array of ZnO microrods was partially covered with an insulator layer of photoresist using spin coating. The thickness of the photoresist film was monitored by microinterferometer MI-4 and was found to be $450 \pm 50$ nm. This stage was followed by deposition of In thin film as the top electrode using the magnetron sputtering method [6]. The In contact was deposited through the shadow mask. A liquid photo-positive resist "Cramolin Positiv Resist" was used as the photoresist-insulator [4, 7]. The schematic image of the photovoltaic cell is shown in Fig. 2.

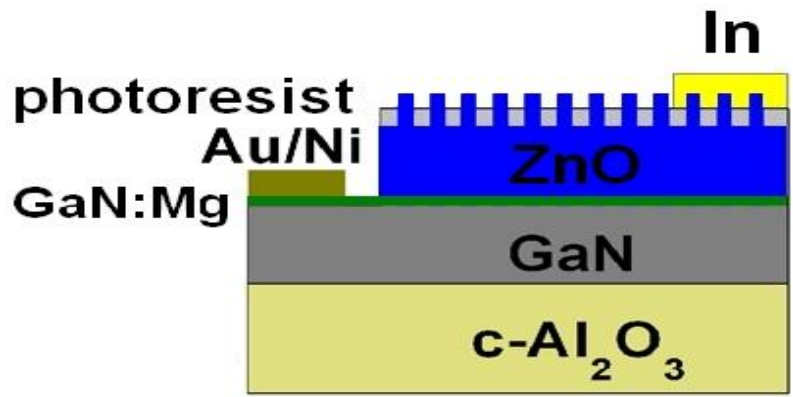

**Fig. 2** Schematic diagram of the *p*-GaN film/ZnO microrods array heterojunction photovoltaic cell structure

Morphology of the sample surface was examined using REMMA-102-02 Scanning Electron Microscope-Analyzer (JCS SELMI, Sumy, Ukraine).

The measurements of the current-voltage (*I-V*) characteristics and photovoltaic properties were carried out using Keithley Model 6514 System Programmable Electrometer (Keithley Instruments Inc., Ohio, USA).

A UV-LED was used as the light source for UV illumination of a photovoltaic cell. It produced a non-polarized exciting radiation with the wavelength of 395 nm, the bandwidth (FWHM) of 13 nm and the power density of the UV light 2 mW/cm$^2$.

The absorption spectra of the obtained samples in the ultraviolet and visible regions were investigated using a portable fiber optic spectrometer AvaSpec-ULS2048L-USB2-UA-RS (Avantes BV, Apeldoorn, Netherlands) with an input slit of 25 μm, a diffraction grating of 300 lines/mm and a resolution of 1.2 nm. A balanced compact deuterium-halogen light source Avantes AvaLight-DHc (200–2500 nm) was used. The detection of light in the spectrometer was carried out by a 2048 pixel CCD detector. The special software AvaSoft 8 (Apeldoorn, Netherlands) for automated computer control for this type of spectrometers and spectra processing was used.

### 3 Results and discussion

According to scanning electron microscopy data, ZnO layer grown on the epitaxial GaN film consists of the tightly packed vertical microrods with a diameter of approximately 2–3 μm (Fig. 1).

The *I-V* characteristics of the created photovoltaic cell with the In/ZnO/GaN:Mg/Ni/Au structure are shown in Fig. 3, 4. The *I-V* curve clearly shows a nonlinear increase of current under the forward bias, indicating reasonable *p–n* junction characteristics.

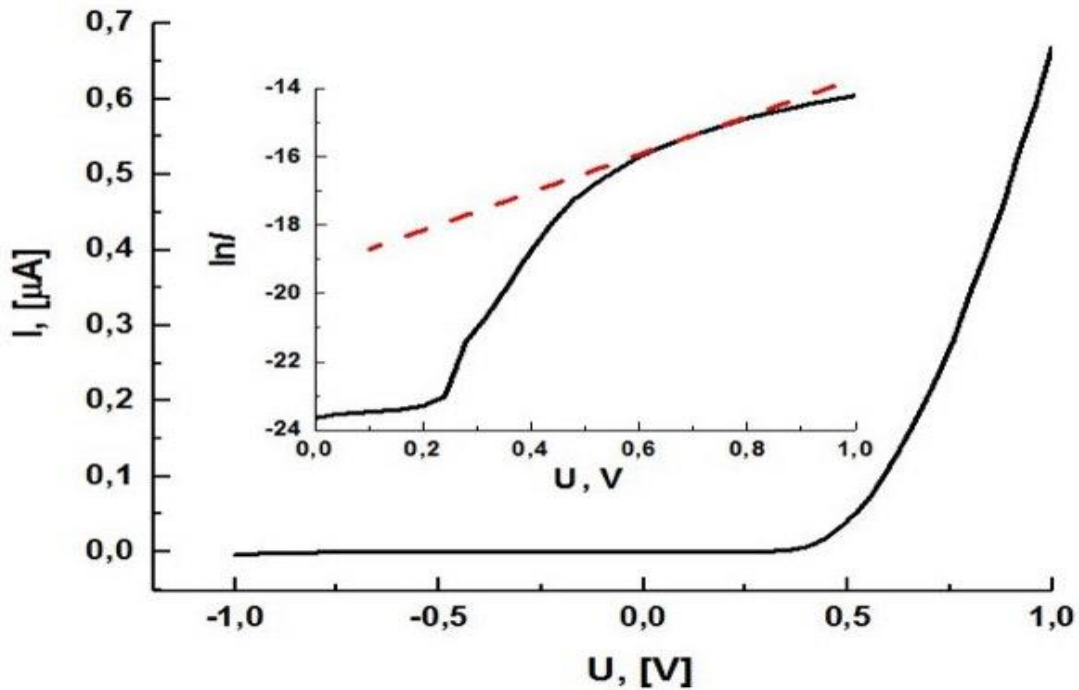

**Fig. 3** *I-V* characteristics of the created photovoltaic cell with In/ZnO/GaN:Mg/Ni/Au structure in darkness. The inset shows the ln(*I*) versus *V* plot to acquire the diode-ideality factor from the slope of the fitting curve.

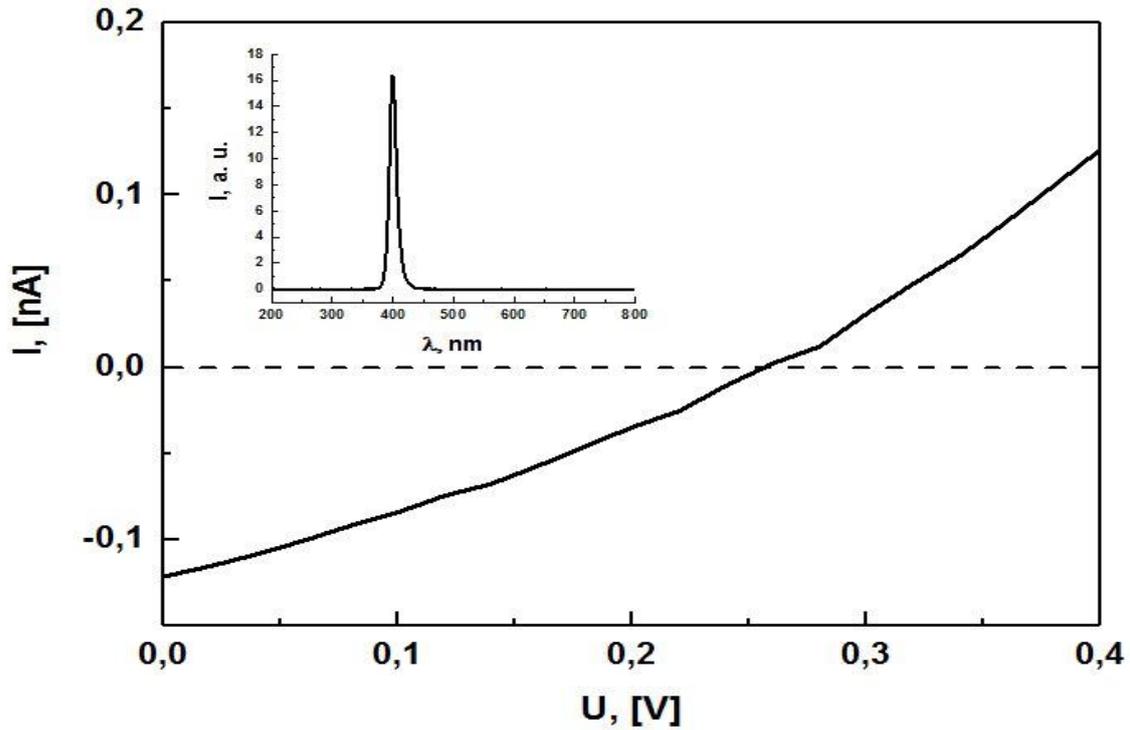

**Fig. 4** *I-V* characteristics of the created photovoltaic cell with In/ZnO/GaN:Mg/Ni/Au structure under UV LED illumination. The inset shows the UV LED emission spectrum.

The rectification coefficient, determined for the voltage of 1 V, was found to be 170. The turn-on voltage of ZnO/GaN (rods/film) heterojunction is around 0.6 V. The diode-ideality factor ($\eta$) was determined from the slope of the ln(*I*) versus *V* plot in the range from 0.6 V to 0.8 V (inset to Fig. 3) using the equation:

$$\eta = e/k_B T \, [\delta(\ln I)/\delta V]^{-1}, \qquad (1)$$

where $k_B$ is the Boltzmann constant and *T* is the operating temperature.

In our case the diode-ideality factor was calculated to be of around 4. This result is in agreement with the data of the current–voltage behavior of *n*-ZnO/*p*-GaN heterostructures presented in [8, 9]. In our previous works, devoted to LEDs based on *n*-ZnO micro- and nanostructures grown by various methods and *p*-GaN films, we obtained values of the ideality factor in the range of 30–45 [3, 4]. The large values of the ideality factor indicate a high density of trap states [10]. The considerable deviation of $\eta$ from the ideal case ($\eta = 1$ for the thermionic emission model) may be also connected with the quality of the contacts to the *p-n* junction [11, 12].

Figure 4 shows the *I-V* curve of the *p*-GaN/*n*-ZnO microrods structure under illumination of UV LED (2 mW/cm$^2$). The *I-V* characteristic of *p*-GaN/*n*-ZnO structures exhibits an open-circuit voltage of 0.26 V, a short-circuit current of 0.124 nA, and a fill factor of 39 % that give the overall efficiency of 1.4·10$^{-5}$ %. It is necessary to note that the obtained value of efficiency is quite low. For comparison, as it was reported in papers [2, 5], the values of efficiency of the photovoltaic cells on the basis of *p*-GaN epitaxial films and *n*-ZnO films measured under simulated AM 1.5 illumination with and without the ZWB2 filter and under 1-Sun illumination were found to be 0.025 %, 0.46 % and 0.001 %, respectively. However, contrary to the cases described in [2, 5], we used for the photovoltaic studies a source of UV light emitting in a much narrower spectral range.

Figure 5 presents the room temperature absorption spectrum of the photoresist film of 450 nm thickness. The spectrum reveals the absorption bands with the maxima at 339, 408 and 602 nm, that is consistent with the technical data sheet of "Cramolin Positiv Resist".

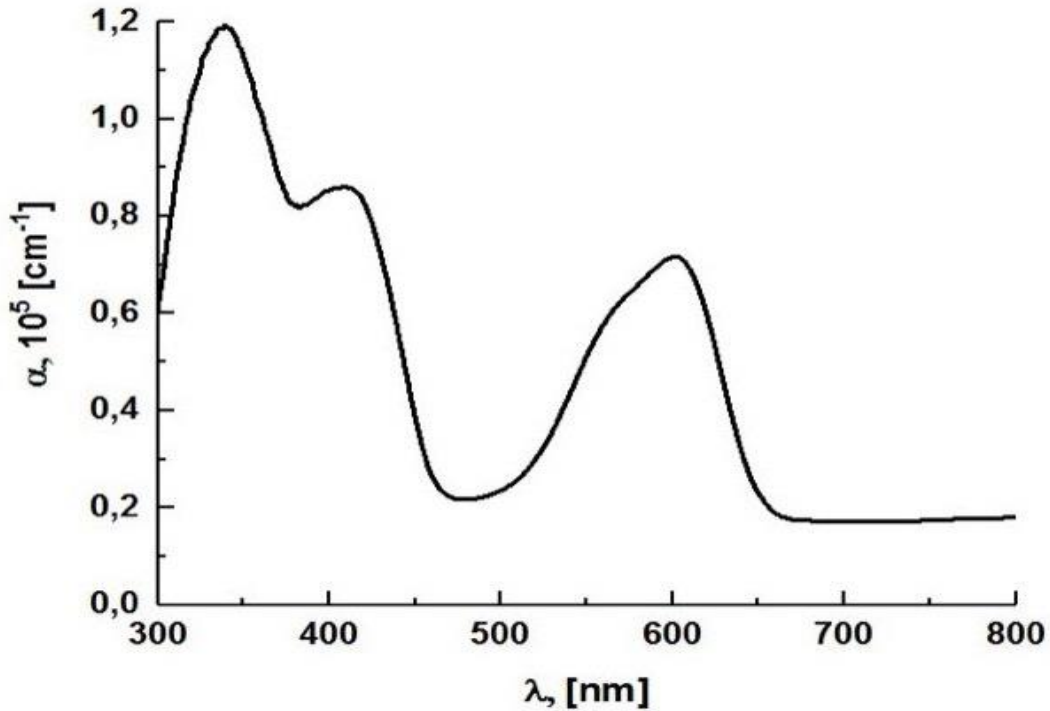

**Fig. 5** The room temperature absorption spectrum of a 0.45 μm thick photoresist film.

On the basis of the literature data [2, 5] and our experimental results one can conclude that the performance of the produced device is mainly limited by the defects at the ZnO/GaN interface and partial light absorption by the photoresist layer. Under such circumstances, the device performance may be improved by reducing of the interface defects and replacing of the chosen photoresist with an UV transparent insulator, such as a $SiO_2$ layer.

**4 Conclusion**

In summary, on can conclude that *n*-ZnO microrods quasiarray/*p*-GaN film photovoltaic cell has been created with the hexagonal ZnO microrods prepared by the method of gas-transport reactions. Under the illumination of UV LED light a clear photovoltaic effect was observed. The fabricated photovoltaic cell device shows a turn-on voltage of 0.6 V. The power conversion efficiency of the photovoltaic cell is $1.4 \cdot 10^{-5}$ % under UV LED illumination. Although this parameter was found to be quite low, the proposed comparatively simple technology and scheme of a solar cell looks as a very suitable basis for further improvements and application. Performed analysis shows the ways of considerable enhancement of the cell performance consisting in reducing of the interface defects and optimization of the photoresist layer parameters. Such an approach would be fruitful for designing of the transparent ultraviolet photovoltaic cells with optimized parameters.

**Acknowledgments**
This work was supported by the Ministry of Education and Science of Ukraine.